\documentclass[3p,twocolumn]{elsarticle}

\usepackage{amsmath}
\usepackage{amsfonts}
\usepackage{colonequals}
\usepackage{epsfig}
\usepackage{url}

\renewcommand{\d}{\mathrm{d}}

\newcommand{\p}{\partial}
\newcommand{\pd}[2]{\frac{\p{#1}}{\p{#2}}}
\newcommand{\pl}[2]{\p{#1}/\p{#2}}

\newcommand{\abs}[1]{\lvert#1\rvert}

\newcommand{\subst}[2]{\left.{#1}\right\vert_{#2}}

\newcommand{\tpa}{\tau+\alpha}
\newcommand{\tpg}{\tau+\gamma}

\newcommand{\lmn}{\lambda-\nu}

\begin{document}

\title{Canonical methods of constructing invariant tori by phase-space sampling}
\author[tut]{Teemu Laakso\corref{cor}}
\ead{teemu.laakso@tut.fi}
\author[tut]{Mikko Kaasalainen}
\address[tut]{Department of Mathematics, Tampere University of Technology, PO Box 553, 33101 Tampere, Finland}
\cortext[cor]{Corresponding author}
\date{\today}

\begin{abstract}
Invariant tori in phase space can be constructed via a nonperturbative canonical transformation applied to a known integrable Hamiltonian $\mathcal{H}$. Hitherto, this process has been carried through with $\mathcal{H}$ corresponding to the isochrone potential and the harmonic oscillator. In this paper, we expand the applicability regime of the torus construction method by demonstrating that $\mathcal{H}$ can be based on a St{\"a}ckel potential, the most general known form of an integrable potential. Also, we present a simple scheme, based on phase space sampling, for recovering the angle variables on the constructed torus. Numerical examples involving axisymmetric galactic potentials are given.
\end{abstract}

\begin{keyword}
near integrability, invariant torus, torus construction, galaxies, Stackel potential
\MSC[2010]  70H07, 70H15, 70K43, 65P10, 85A05
\end{keyword}











\maketitle

\section{Introduction}
Poincar\'e called the problem of determining the dynamical behaviour of a system given by a perturbed integrable Hamiltonian ``the fundamental problem of dynamics''. He was quite right in doing so, as the problem has indeed proven to be one of the cornerstones of modern dynamics. The inverse version of this problem, though just as interesting, has attracted less attention. One reason for this is that, as befits an inverse problem, it is quite challenging and without a unique solution. Both problems can be expressed by the fundamental equation
\begin{equation*}
H(\theta,J)=H_0(J)+\epsilon H_1(\theta,J),
\end{equation*}
where $H_0:\mathbb{R}^n\to\mathbb{R}$ is an integrable Hamiltonian, $H,H_1:\mathbb{R}^{2n}\to\mathbb{R}$ are general Hamiltonians, $\theta\in\mathbb{R}^n$, $J\in\mathbb{R}^n$ are action-angle variables corresponding to $H_0$, and $\epsilon$ is a (small) constant. In the direct problem, $H_0$ and $\epsilon H_1$ are known, while in the inverse one we know an $H$ in the form $H(p,q)$, where $p\in\mathbb{R}^n$, $q\in\mathbb{R}^n$ are some canonically conjugate coordinates, and want to find an $H_0$ and its $J$ (and the corresponding $\theta$) such that $\epsilon H_1$ is as small as possible (in some sense). In other words, we want to find the integrable Hamiltonian $H_0$ best approximating a given, typically near-integrable, Hamiltonian $H$. The key to this problem lies in the explicit construction of the suitable phase-space tori, labelled by $J$, with which $H_0$ can be defined \citep{KB1994b,Pos1982}.

The general problem of torus construction must be approached nonperturbatively, as by definition we do not know any tori that could be perturbed KAM-like into some other suitable invariant tori. Such a route was taken by \citet{MB1990}, who presented a method (canonical torus modelling), where a model torus is defined by applying a canonical transformation in the form of a Fourier series to an integrable toy Hamiltonian $\cal H$. The model is fitted to a sample of phase-space points in such a way that it reconstructs any existing KAM tori of the given target Hamiltonian $H$, or creates invariant tori of some $H_0$ mimicking the properties of $H$ in phase-space regions where $H$ has no tori of its own. Ideally, $H_0=H$ on all KAM tori of $H$. As shown by \citet{Kaa1994,Kaa1995} the method can be applied even in chaotic or strongly resonant regions of the target phase-space.

In a wider scope, finding invariant structures in near-integrable systems is a problem of theoretical interest, but also encountered in many physical applications. \citet{LCC2006} provide a useful list of references related to the topic. Numerical methods based on Fourier series are used to study particle physics \citep{War1991}, superconductivity \citep{LBH2005}, quantum dynamics \citep{GMB2007}, celestial mechanics \citep{CJ2000}, etc. Often, these kinds of methods are general in nature, and many are also introduced as such \citep{LCC2006, JO2004}.

The canonical torus modelling must be specially tuned for each pair of $\cal H$ and $H$. \citet{MB1990} and \citet{KB1994a,KB1994b} demonstrated the method in gravitational systems by using the analytically integrable potentials of the harmonic oscillator and the isochrone for $\cal H$. Here we present a natural continuation of this approach by employing St\"ackel potentials, the widest category of known integrable systems whose Hamilton-Jacobi equation is explicitly separable, in our $\cal H$. Another, more application-dependent reason for using St\"ackel potentials is that they produce the same topologically different orbit families that are typically found in large gravitationally bound systems. Thus one does not have to use different $\cal H$ for different orbit types. Moreover, elliptic St\"ackel potentials resemble the potentials of such systems much more than the isochrone or the harmonic oscillator. This reduces the need for additional point transformations needed to twist the orbits inherent to $\cal H$ into suitable shapes as in \citet{KB1994a,KB1994b}. 

The useful and natural properties of St\"ackel potentials make the algorithm low-maintenance, which is desirable in applications such as determining the potential of our Galaxy from astrophysical data \citep{Bin2005}. Besides, we have the possibility of readily producing fully three-dimensional models for gravitational systems (triaxial galaxy models). In terms of computational cost, the advantages introduce some overhead: although separably computable in a closed form, the action-angle variables in a nontrivial St{\"a}ckel potential do not have analytical expressions, which has an impact on the complexity of the implementation.

As a step towards the goals above, we have developed the analytical and computational tools for constructing invariant tori from an axisymmetric ellipsoidal St{\"a}ckel toy Hamiltonian $\cal H$. A solid basis for such development is given by \citet{Zee1985}. He introduces the perfect ellipsoid, a triaxial galaxy model which produces a St{\"a}ckel potential, and recreates the major orbit families that \citet{Sch1979} found by numerically simulating an elliptical galaxy in dynamical equilibrium. In this paper, we implement $H$ as an axisymmetric special case, the perfect oblate spheroid. This galaxy model was first discovered by \citet{Kuz1956}. There is only one orbit family, the short-axis tube. In a specific numerical example, we construct invariant tori of an $H_0$ approximating the Hamiltonian $H$ of the axisymmetric logarithmic potential.

A major benefit of torus modelling is that one can create a system of action-angle variables for $H_0$. The angle variables on the model torus are not needed during the construction process. However, they can be recovered afterwards by various means:  \citet{BK1993} derived a partial differential equation for the model angles, and solved it using discrete Fourier transforms. \citet{KB1994a} used a method based on orbit integration, and found it superior. However, integrated orbits may quickly stray off the model torus, if it happens to be in a chaotic region of $H$. Here we introduce a general technique of computing the model angles. It is simple and based on the same phase-space sampling as in the torus algorithm itself.

The paper is organized as follows. First, we outline the torus modelling algorithm, and the new method of obtaining the model angles. Next, we review the process of computing the action-angle variables of the perfect oblate spheroid, and discuss some aspects in the implementation. A numerical example follows, where we demonstrate the presented methods.

\section{Torus modelling}
Let ${\cal H}:\mathbb{R}^3\times\mathbb{R}^3\to\mathbb{R}$ be an integrable Hamiltonian. Suppose that the corresponding Hamilton-Jacobi equation separates in some canonical phase-space coordinates $(q,p)$. It follows that we can explicitly solve three integrals of motion $I\in\mathbb{R}^3$. These restrict the motion in the phase space to a subset $\mathcal{A}\subset\mathbb{R}^3\times\mathbb{R}^3$ which, for periodic motion, is homeomorphic to a torus $\mathbb{T}^3$. The natural coordinates on the torus are the action-angle variables $(\vartheta,\mathcal{J})$. We have a coordinate transformation $\mathcal{A}\to\mathcal{A}$,
\begin{equation}
\label{e:map:toy}
(q,p)\leftrightarrow(\vartheta,\mathcal{J}).
\end{equation}

Consider a canonical transformation, given by the generating function
\begin{equation*}
\label{e:def:F}
F(\vartheta,J)=\vartheta\cdot J-i\sum_{k\in D}S_k(J)\exp(ik\cdot\vartheta),
\end{equation*}
where $D\subset\mathbb{Z}^3\backslash\{0\}$ is a set of multi-indices (wave numbers), and $S_k\in\mathbb{C}$ are (Fourier) coefficients. Explicitly,
\begin{align}
\label{e:trans:J}
\mathcal{J}&=J+\sum_{k\in D}kS_k(J)\exp(ik\cdot\vartheta),\\
\label{e:trans:theta}
\theta&=\vartheta-i\sum_{k\in D}\pd{S_k(J)}{J}\exp(ik\cdot\vartheta).
\end{align}
The map $\mathcal{A}\to\mathcal{A}$,
\begin{equation*}
\label{e:map:fourier}
(\vartheta,\mathcal{J})\leftrightarrow(\theta,J),
\end{equation*}
establishes new phase space coordinates $(\theta,J)$ which we identify as action-angle variables; not those of $\mathcal{H}$, but some other integrable Hamiltonian $H_0$.

Let $\mathcal{T}\subset\mathbb{R}^3\times\mathbb{R}^3$ be a subset of phase space where $J$ is constant. $\mathcal{T}$ is homeomorphic to $\mathbb{T}^3$, and we call it the model torus. Through the coordinate transformation $\mathcal{T}\to\mathcal{T}$,
\begin{equation}
\label{e:map:torus}
(\theta,J)\leftrightarrow(q,p),
\end{equation}
we define $(\theta,J)$ as action-angle variables of an integrable Hamiltonian $H_0$ which restricts the motion to $\mathcal{T}$.

In torus modelling, the idea is to choose the coefficients $S_k$, $k\in D$ in such a way that the model torus $\mathcal{T}$ becomes an invariant torus of an
$H_0$ close to the given target Hamiltonian $H:\mathbb{R}^3\times\mathbb{R}^3\to\mathbb{R}$ and identical to it on its existing KAM tori. This is achieved by studying $H$ through the map $\mathcal{T}\to\mathbb{R}:$
\begin{equation*}
(\vartheta,J)\mapsto(\vartheta,\mathcal{J})\mapsto(q,p)\mapsto H(q,p)
\end{equation*}
which is independent of $\pl{S_k}{J}$ (the coefficients in Eq.\ \eqref{e:trans:theta}). An algorithm is introduced, as follows. First, we choose the value of $J$, and a grid of points $\vartheta_{(m)}$, $m=1,\ldots,M$ in the toy angle space.  On the grid, we minimize
\begin{equation}
\label{e:def:chi2}
\chi^2=\sum_{m=1}^M\left[H(\vartheta_{(m)},J)-\bar{H}\right]^2,
\end{equation}
where $\bar{H}$ is the arithmetic mean of $H$ over the grid;
\begin{equation*}
\bar{H}=\frac{1}{M}\sum_{m=1}^MH(\vartheta_{(m)},J).
\end{equation*}
After the minimization we have, on the model torus, $H(\vartheta,J)\approx H_0(J)\colonequals\bar{H}$, which means that the model torus $\mathcal{T}$, defined by $J$ and $S_k(J)$, $k\in D$, either approximates an existing invariant torus of $H$ or creates a close equivalent in some region of phase space.

The domain of the angle variables $\vartheta$ is trivially $[0,2\pi)^3$. The image of the map $(\vartheta,J)\leftrightarrow(q,p)$ is the model torus $\mathcal{T}$. Hence, at this point, we can access $\mathcal{T}$ as a geometrical object in the phase space without using \eqref{e:trans:theta}. We can, e.g., plot its Poincar{\'e} sections.

\section{Model angles by phase-space sampling}
The model angles $\theta$ were not needed in the torus algorithm above. However, in order to complete the set of variables $(\theta,J)$ on the model torus $\mathcal{T}$, and to use the map \eqref{e:map:torus} explicitly, one needs to find values for the coefficients $\pl{S_k}{J}$.

Suppose that we have found coefficients $S_k$ in such a way that $J$ corresponds to a suitable invariant torus of $H_0\approx H$. On this torus, the model frequencies are
\begin{equation*}
\omega=\pd{H}{J}.
\end{equation*}
By denoting $u\colonequals(q,p)$, $u\in\mathbb{R}^3\times\mathbb{R}^3$, we have, for a fixed value of $\vartheta$,
\begin{equation}
\label{e:omega:model}
\omega=\pd{H}{u}\pd{u}{\mathcal{J}}\pd{\mathcal{J}}{J},
\end{equation}
where $\pl{u}{\mathcal{J}}$ is a $6\times 3$ matrix, $\pl{\mathcal{J}}{J}$ has $3\times 3$ elements:
\begin{equation*}
\label{e:pd:J:J:model}
\pd{\mathcal{J}}{J}=I+\sum_{k\in D} \left[k\pd{S_k}{J}\right]\exp(ik\cdot\vartheta),
\end{equation*}
and $I$ is a $3\times 3$ identity matrix. The $3\times 3$ 
matrix $[k\pl{S_k}{J}]$ is formed by the outer product of $k$ and 
$\pl{S_k}{J}$:
\begin{equation*}
\left[k\pd{S_k}{J}\right]_{ij}=k_i\, \pd{S_k}{J_j}.
\end{equation*}
Substituting, and considering each component $\omega_n$, $n=1,2,3$ separately, we have
\begin{equation*}
\omega_n-\pd{H}{u}\pd{u}{\mathcal{J}}\sum_{k\in D} k\pd{S_k}{J_n}\exp(ik\cdot\vartheta)=\pd{H}{u}\pd{u}{\mathcal{J}_n}
\end{equation*}
which is a linear equation for the variables $\omega_n$ and $\pl{S_k}{J_n}$, $k\in D$. Since
\begin{equation}
\label{e:pd:H:Sk}
\pd{H}{S_k}=\pd{H}{u}\pd{u}{\mathcal{J}}k\exp{(ik\cdot\vartheta)},
\end{equation}
we can also write this as
\begin{equation}
\label{e:def:angle:linear}
\omega_n-\sum_{k\in D}\pd{H}{S_k}\pd{S_k}{J_n}=\pd{H}{u}\pd{u}{\mathcal{J}_n}.
\end{equation}

Let $S\in\mathbb{C}^{\#D}$ be a vector which contains all of the coefficients $S_k$, $k\in D$. By dictating that Eq.\ \eqref{e:def:angle:linear} should hold on a grid of angles $\vartheta_{(m)}$, $m=1,\ldots,M$, we obtain a linear system $X\beta=y$, where each row $X_m$ of $X$ has the length $\#D+1$;
\begin{equation*}
X_m=
\begin{bmatrix}
1 & -\displaystyle\subst{\pd{H}{S}}{J,\vartheta_{(m)}}
\end{bmatrix},
\end{equation*}
\begin{equation*}
\beta=
\begin{bmatrix}
\omega_n & \displaystyle\pd{S}{J_n},
\end{bmatrix}^T,
\end{equation*}
and the components $y_m$ of $y$ are
\begin{equation*}
y_m=\subst{\pd{H}{u}\pd{u}{\mathcal{J}_n}}{J,\vartheta_{(m)}}
\end{equation*}
The least-squares solution for $\beta$ is given by the normal equations
\begin{equation}
\label{e:def:angle:normal}
X^TX\beta=X^Ty.
\end{equation}
The defined values of $H_0$ and $\omega$ on the constructed tori are self-consistent since $H_0\equiv\bar{H}\implies\omega=\pl{H_0}{J}=\overline{\pl{H}{J}}$.

\section{Oblate spheroidal St{\"a}ckel potentials}
Ellipsoidal systems are most naturally represented in ellipsoidal coordinates. Adopting the notation of \citet{Zee1985}, we define the triaxial ellipsoidal coordinates as the roots $\tau=\lambda,\mu,\nu$ of 
\begin{equation*}
\frac{x^2}{\tpa}+\frac{y^2}{\tau+\beta}+\frac{z^2}{\tpg}=1,
\end{equation*}
where $x,y,z$ are the Cartesian coordinates in $\mathbb{R}^3$, and $\alpha,\beta,\gamma\in\mathbb{R}$ are coordinate parameters. Suppose that the parameters are selected in such a way that the coordinate surfaces match the geometry of the modelled system. When the model changes from triaxial to oblate spheroidal (axisymmetric with respect to $z$-axis), coordinates become the prolate spheroidal coordinates $\lambda,\phi,\nu$. Two of these are the roots $\tau=\lambda,\nu$ of
\begin{equation*}
\label{e:def:elliptical}
\frac{R^2}{\tpa}+\frac{z^2}{\tpg}=1,
\end{equation*}
where $R^2=x^2+y^2$. The coordinates $\lambda,\nu$ are elliptical coordinates in the meridional plane the orientation of which is labelled by the azimuthal angle $\phi$. The prolate spheroidal coordinates can also be represented by trigonometric functions, but in order to ease the future transition to fully triaxial models, we adopt the definition above.

\begin{figure}[htbp]
\includegraphics[width=\columnwidth]{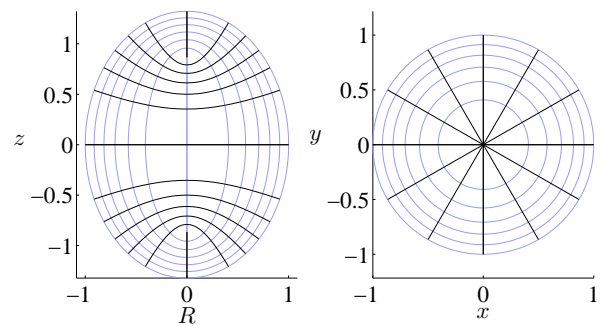}
\caption{(Colour online) Prolate spheroidal coordinate surfaces ($\alpha=-1$, $\gamma=-0.25$) in the meridional (left) and equatorial (right) planes.}
\label{f:coord:surf}
\end{figure}

If we set $\alpha<\gamma$, and choose $\lambda>\nu$, we have $(\lambda,\nu)\in[-\alpha,\infty)\times[-\gamma,-\alpha]$. We select $\phi\in[0,\pi/2)$, and introduce an additional discrete variable $n\in\{-1,1\}^3$ which identifies the octants in $\mathbb{R}^3$. Let $R,\varphi,z$ be the standard cylindrical coordinates. For the set of points $\mathbb{R}^3\backslash\{R=0\}$ we have a one-to-one correspondence:
\begin{equation*}
\label{e:map:coordinates}
(x,y,z)\leftrightarrow(R,\varphi,z)\leftrightarrow(\lambda,\phi,\nu,n).
\end{equation*}
An example of the prolate spheroidal coordinate surfaces is shown in Fig.\ \ref{f:coord:surf}. Surfaces of constant $\lambda$ and $\nu$ are axisymmetric ellipsoids and hyperboloids, respectively.

Consider an axisymmetric potential $\Psi:\mathbb{R}^2\to\mathbb{R}$, $(\lambda,\nu)\mapsto\Psi(\lambda,\nu)$ in the prolate spheroidal coordinates. The Hamiltonian ${\cal H}:\mathbb{R}^2\times\mathbb{R}^3\to\mathbb{R}$ which describes the motion in $\Psi$ satisfies ${\cal H}(\lambda,\nu,p_\lambda,p_\phi,p_\nu)=E$, where $p_\tau$ are the conjugate momenta to $\tau=\lambda,\phi,\nu$, and $E$ is the constant energy (per unit mass). Due to axisymmetricity, the $z$-component of the angular momentum is also constant, and our system could be reduced to planar. However, again, for the sake of future expandability, we treat the model as three-dimensional.

The Hamilton-Jacobi equation for $\cal H$ is
\begin{equation}
\label{e:def:HJE}
\sum_\tau\frac{1}{2h_\tau^2}\left(\pd{W}{\tau}\right)^2+\Psi-E=0,
\end{equation}
where the sum is taken over $\tau=\lambda,\phi,\nu$, and $h_\tau$ are the scale factors of the prolate spheroidal coordinates. The unknown function $W:\mathbb{R}^3\to\mathbb{R}$, $(\lambda,\phi,\nu)\mapsto W(\lambda,\phi,\nu)$ is Hamilton's characteristic function.

By definition, the Hamilton-Jacobi equation for a St{\"a}ckel potential is separable. In prolate spheroidal coordinates such a potential can be written as
\begin{equation}
\label{e:def:stackel}
\Psi(\lambda,\nu)=-\frac{f_\lambda(\lambda)-f_\nu(\nu)}{\lmn},
\end{equation}
where $f_\lambda$ and $f_\nu$ are arbitrary smooth real-valued functions. Solving Eq.\ \eqref{e:def:HJE} with \eqref{e:def:stackel} yields
\begin{equation*}
\label{e:solution:HJE}
W(\lambda,\phi,\nu)=\sum_\tau W_\tau(\tau),
\end{equation*}
where $W_\tau:\mathbb{R}\to\mathbb{R}$, and $\tau=\lambda,\phi,\nu$. There are two independent separation constants which, in addition to the total energy, translate to three integrals of motion; $\cal H$, $I_2$, and $I_3$.

If we interpret the constants $I=({\cal H},I_2,I_3)$ as new canonical momenta, $W(\lambda,\phi,\nu,{\cal H},I_2,I_3)$ acts as a generating function which defines a canonical transformation; we have $p_\tau=\pl{W}{\tau}$ for $\tau=\lambda,\phi,\nu$, and the conjugate coordinates to $I$ are $\pl{W}{I}$.

Suppose that the motion is quasiperiodic. The action integrals $\mathcal{J}=(\mathcal{J}_\lambda,\mathcal{J}_\phi,\mathcal{J}_\nu)$ are defined as
\begin{equation*}
\label{e:def:Jtau:general}
\mathcal{J}_\tau=\frac{1}{2\pi}\oint p_\tau(\tau',I)\d\tau',
\end{equation*}
where the closed path integral is taken over a complete period of motion in the corresponding direction $\tau=\lambda,\phi,\nu$. The conjugate angles $\vartheta=(\vartheta_\lambda,\vartheta_\phi,\vartheta_\nu)$ are
\begin{equation*}
\label{e:def:thetatau}
\vartheta_\tau=\pd{W}{\mathcal{J}_\tau}=\pd{W}{I}\pd{I}{\mathcal{J}_\tau},
\end{equation*}
for each $\tau=\lambda,\phi,\nu$.

\section{Algorithmic details}
A detailed description, including explicit formulas and derivations, of using oblate spheroidal St{\"a}ckel toy Hamiltonians in torus modelling is given in \citet{Laa2011}. Here, we give an overview, and focus on some essential points. 

For an oblate spheroidal St{\"a}ckel Hamiltonian, equations above outline the explicit steps in the coordinate transformation \eqref{e:map:toy}, from left to right. The integrals of motion $I$ can be solved algebraically, but numerical methods are required in various other places. First of all, we need one-dimensional root-finding algorithms in order to define the boundaries of the set $\mathcal{T}$ in the prolate spheroidal coordinates ($p_\tau^2(\tau,I)\ge 0$, $\tau=\lambda,\nu$). For $\pl{W}{I}$ and $\mathcal{J}$ we need to solve partly improper integrals with numerical quadrature rules. When \eqref{e:map:toy} is implemented from right to left, $(\lambda,\phi,\nu,n)$ and $I$ are evaluated using multi-dimensional root-finding algorithms. 

Although the prolate spheroidal coordinates $\tau=\lambda,\phi,\nu$ are suitable for numerical work, their conjugate momenta $p_\tau$ (or coordinate velocities $\dot\tau$) are not; points in the equatorial plane $\nu=-\gamma$ are singular in coordinate transformations, and in partial derivatives with respect to $p_\nu$. Fortunately, these singularities can be avoided by manipulating equations with the polar coordinates $u:=(R,\varphi,z,p_R,p_\varphi,p_z)$  as the frame of reference in phase space.

Additional numerical quadratures are needed in torus modelling, if we minimize the r.h.s.\ of Eq.\ \eqref{e:def:chi2} using gradient-based methods. We have to evaluate the partial derivatives $\pl{H}{S_k}$, and especially the matrix $\pl{u}{\mathcal{J}}$, as Eq.\ \eqref{e:pd:H:Sk} suggests. The same partial derivatives are also needed when solving the model angles from Eq.\ \eqref{e:def:angle:linear}. We obtain $\pl{u}{\mathcal{J}}$ by inverting $\pl{(\vartheta,\mathcal{J})}{u}$. Especially the computation of $\pl{\vartheta}{u}$ is a nontrivial exercise.

Terms in the Fourier series \eqref{e:trans:J} and \eqref{e:trans:theta} can be simplified or cancelled out, if the target Hamiltonian $H$ shares some of the symmetries of $H$. We shall use a target which is time reversible, and symmetric about $R=0$ and $z=0$. This implies that each $S_k(J)=S_k(J_\lambda,J_\nu)\in\mathbb{R}$. The transformations \eqref{e:trans:J} and \eqref{e:trans:theta} become
\begin{align}
\label{e:trans:J:pos}
\mathcal{J}&=J+2\sum_{k\in D_{\lambda\nu}^+}kS_k(J_\lambda,J_\nu)\cos(k\cdot\vartheta),\\
\label{e:trans:theta:pos}
\theta&=\vartheta+2\sum_{k\in D_{\lambda\nu}^+}\pd{S_k(J_\lambda,J_\nu)}{J}\sin(k\cdot\vartheta),
\end{align}
where the set $D_{\lambda\nu}^+$ is obtained from $D_{\lambda\nu}=\{(k_\lambda,k_\phi,k_\nu)\in\mathbb{Z}^3\backslash\{0\}:k_\phi=0\}$ by removing one of the two elements $k\in D_{\lambda\nu}$ for which $k=-k$. We have $\mathcal{J}_\phi=J_\phi$ and $\theta_\phi=\vartheta_\phi$.

\section{Numerical examples}
We use the perfect oblate spheroid as the toy potential in the torus algorithm. The arbitrary functions in the St{\"a}ckel potential \eqref{e:def:stackel} are thus
\begin{equation*}
f_\tau(\tau)=-2\pi G \rho_0\alpha\sqrt{-\gamma(\tpg)}\arctan\sqrt{\frac{\tpg}{-\gamma}},
\end{equation*}
for $\tau=\lambda,\nu$. $G$ is the gravitational constant, and $\rho_0$ is the density at the origin. We set $G=1$.

For the target Hamiltonian, we use the logarithmic potential in the meridional plane:
\begin{equation*}
\Phi(R,z)=\frac{1}{2}\ln\left(R^2+\frac{z^2}{a^2}+b^2\right),
\end{equation*}
where $a=0.8$ and $b=0.14$. Thus $H=\abs{p}^2/2+\Phi$, where $p$ is the momentum vector.
 
Before running the actual torus modelling algorithm, values for the parameters $\alpha$, $\gamma$, and $\rho_0$ are chosen in such a way that the toy and target potential surfaces are as similar as possible in configuration space. We formulate this as a curve fitting problem which we solve in a sparse grid of points using the Levenberg-Marquardt algorithm. As a result, we have $\alpha=-0.639$, $\gamma=-0.142$, and $\rho_0=1.29$.

As an example, we ran the torus modelling algorithm for $J=(0.5,0.45,0.5)\equalscolon J_0$. The included number of Fourier coefficients $S_k$, $k\in D_{\lambda,\nu}^+$ was determined by trial and error. We found out that reasonable accuracy was achieved by including terms for which $|k_\lambda|\le 96$ and $|k_\nu|\le 24$. The toy angle grid was chosen to contain $100\times 100$ points, which exceeds the Nyquist rate of the Fourier series.

\begin{figure}[hbtp]
\includegraphics[width=\columnwidth]{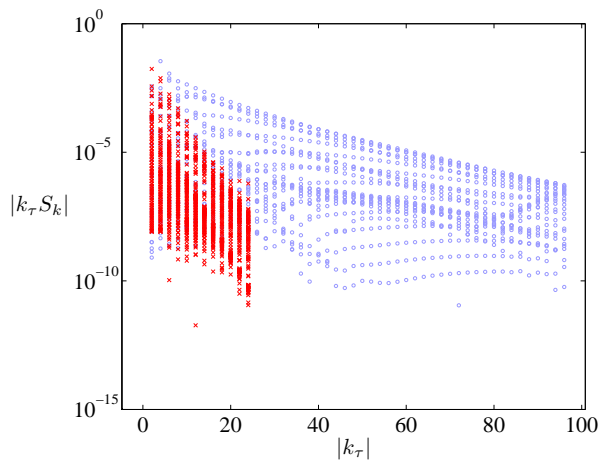}
\caption{(Colour online) Magnitudes of the torus coefficients. Terms $|k_\lambda S_k|$ are plotted with light blue circles, and the terms $|k_\nu S_k|$ with red crosses.}
\label{f:coef}
\end{figure}

The $\chi^2$ minimization with Levenberg-Marquardt converged within a few steps; the magnitudes of the resulting Fourier coefficients are displayed in Fig.\ \ref{f:coef}.
In this case, the Fourier series converges for $\mathcal{J}_\lambda$ about five times more slowly than for $\mathcal{J}_\nu$.

\begin{figure}[htbp]
\includegraphics[width=\columnwidth]{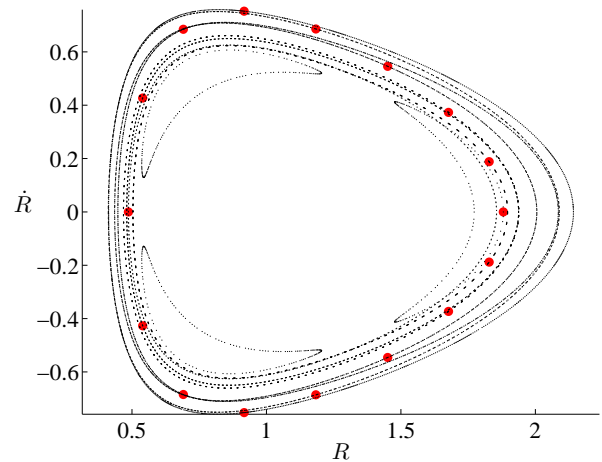}
\caption{(Colour online) Initial ($S_k=0$, $\forall k$) Poincar{\'e} section $z=0$, $p_z>0$ for the model torus (big red dots), and for superimposed trajectories in the target Hamiltonian $H$ (small black dots).}
\label{f:torus:orig}
\end{figure}

\begin{figure}[htbp]
\includegraphics[width=\columnwidth]{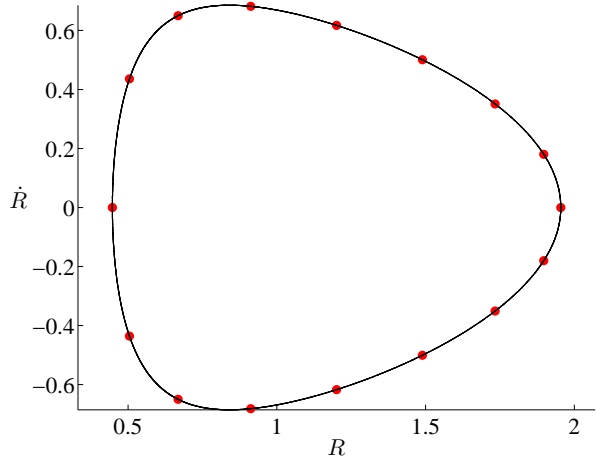}
\caption{(Colour online) Final Poincar{\'e} section with optimized Fourier coefficients (cf. Fig.\ \ref{f:torus:orig})}
\label{f:torus:fitted}
\end{figure}

Since axisymmetric systems are effectively two-dimensional, we may evaluate our success in torus modelling by plotting Poincar{\'e} sections. In Fig.\ \ref{f:torus:orig} we have the section $z=0$, $p_z>0$, computed with the initial, unoptimized coefficients $S_k=0$, $\forall k$. There are 16 discrete points, computed by choosing a grid in $R$, and solving for $\dot{R}$ on the model torus $J=\mathcal{J}$. From each of these points an orbit in the target Hamiltonian $H$ is integrated, and superimposed in the figure. 
Prior to the $\chi^2$ minimization, the orbits clearly do not stay on the model torus. Figure \ref{f:torus:fitted} shows how the situation changes after optimization.
Within observable accuracy, the integrated orbits now coincide with the model torus $J$.

\begin{figure}[htbp]
\includegraphics[width=\columnwidth]{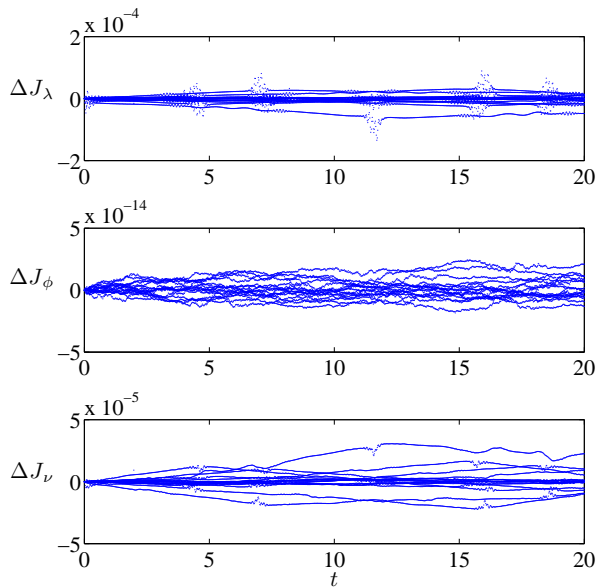}
\caption{(Colour online) Changes in model actions $J$ along target orbits. Each orbit is initially on the model torus $J_0$ (big red dots in Fig.\ \ref{f:torus:fitted}).}
\label{f:actions}
\end{figure}

As a more quantitative indicator of accuracy, we compute the time evolution of $J$ along the integrated target orbits. In Fig.\ \ref{f:actions}, for each orbit, we have plotted the change $\Delta J(t)=J(t)-J_0$, where $J$ is obtained from \eqref{e:trans:J:pos} by approximating $S_k(J)\approx S_k(J_0)$ on the target torus.
The accuracy of the torus model seems consistent with the included coefficients (Fig.\ \ref{f:coef}). A better fit would be achieved by using a longer Fourier series and a denser angle grid.

\begin{figure}[htbp]
\includegraphics[width=\columnwidth]{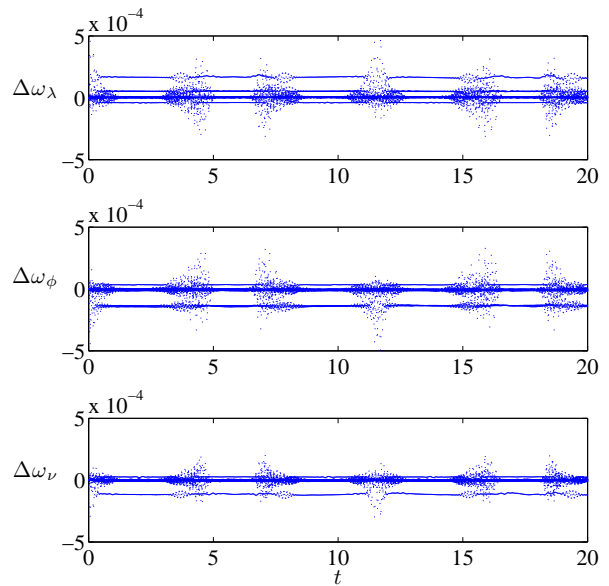}
\caption{(Colour online) Changes in model frequencies $\omega$ along target orbits. Initial values as in Fig.\ \ref{f:actions}.}
\label{f:freq}
\end{figure}

The model angles $\theta$ and frequencies $\omega$ were computed with the phase-space sampling technique. By reusing the toy angle grid from the $\chi^2$ minimization, the matrix $X$ in the normal equations \eqref{e:def:angle:normal} had a size of $1213\times 10000$. Numerical solution for $\beta$ was obtained through LU-decomposition. Validity was checked by computing the change $\Delta \omega(t)=\omega(t)-\omega_0$, where $\omega_0$ corresponds to $J_0$, along the target orbits. This is shown in Fig.\ \ref{f:freq}. The frequencies $\omega$ are computed from \eqref{e:omega:model} using the approximation $\pl{S_k(J)}{J}\approx\pl{S_k(J_0)}{J}$ on the target torus.
Since errors accumulate from two Fourier series, the variation in $\omega$ is somewhat higher than the variation in $J$, but again, results are consistent, and show that, up to a reasonable accuracy, the obtained $(\theta,J)$ act numerically as the action-angle variables of an $H_0$ approximating $H$. The fuzzy parts of trajectories in Fig.\ \ref{f:freq} (and \ref{f:actions}) correspond to phase-space regions, where the $\chi^2$ minimization was least successful. 

The numerical experiments above were repeated for several values of $J$. As a general rule, we found out that the Fourier series converged faster for thinner tori. On the other hand, arbitrarily thin tori could not be mapped by \eqref{e:trans:J} because of the requirement $\mathcal{J}\ge 0$. When probing small values of $J_\lambda$ with $J_\phi=0.45$ and $J_\nu=0.25$, the thinnest mappable torus was around $J_\lambda=0.01$. For tori thicker than $J_0$, more Fourier terms were required for maintaining a similar level of accuracy. In the comfortable region of the target actions $J$, the results were typically as good as in Fig.\ \ref{f:coef}-\ref{f:freq}.

\section{Discussion}
We have demonstrated that an axisymmetric ellipsoidal St{\"a}ckel potential can be used for constructing invariant tori, and associated action-angle variables, for an integrable approximation of an axisymmetric logarithmic potential. Similar targets should yield equally successful results, since the target effectively acts as a plug-in, and can be changed in the torus algorithm with a relatively small effort. The determination of the action-angle variables of the tori is based on phase-space sampling rather than orbit integration, making the procedure manifestly geometric in character.

In the examples above, within a comfortable range of the model actions, the convergence of the algorithm was excellent, but the suitable number of Fourier terms had to be sought by trial and error. We found out that for $\mathcal{J}_\lambda=\mathcal{J}_\nu$, considerably different amounts of terms were required in the $\lambda$ and $\nu$ directions. Using an unbalanced number of Fourier terms is uneconomical, since the number of points in the $\vartheta$-grid is chosen according to the highest frequency in the Fourier series. One possibility is to balance the situation by adjusting the toy potential (and coordinate) parameters during $\chi^2$ optimization as in \citet{MB1990} and \citet{KB1994a}, although this increases the computational overhead somewhat.

The main advantage of modelling with St{\"a}ckel potentials, as opposed to the special analytically integrable ones; isochrone, harmonic oscillator, etc.\ is in their generality. They possess the ability to mimic the full range of orbital features. However, this freedom comes at a cost, since the Hamilton-Jacobi equation must be solved numerically, and the computation times for St{\"a}ckel models are considerably higher. In this paper, we have studied the perfect oblate spheroid which is a special case of the perfect ellipsoid. The success with the axisymmetric St{\"a}ckel potential paves way to future three-dimensional models accessible with a relatively straightforward expansion of our algorithm.

\bibliographystyle{elsarticle-num-names}
\bibliography{teemu.bib}

\end{document}